%% file: main.tex
\documentclass[letterpaper, 10 pt, conference]{ieeeconf}  

\IEEEoverridecommandlockouts                              
\usepackage{graphics} 
\usepackage{epsfig} 
\usepackage{amsmath} 
\usepackage{amssymb}  
\usepackage{physics}
\usepackage{qcircuit}
\usepackage{multirow}
\usepackage[hang,small,bf]{caption}    

\usepackage{tikz}	
\usetikzlibrary{backgrounds,fit,decorations.pathreplacing}  

\usepackage{url}
\usepackage[ruled, vlined, linesnumbered]{algorithm2e}
\usepackage{verbatim} 
\usepackage{soul, color}
\usepackage{lmodern}
\usepackage{fancyhdr}
\usepackage[utf8]{inputenc}
\usepackage{fourier} 
\usepackage{array}
\usepackage{makecell}
\usepackage{subcaption}
\usepackage{tabularx}

\SetNlSty{large}{}{:}

\makeatletter

\newcommand{\Rom}[1]{\expandafter\@slowromancap\romannumeral #1@}
\makeatother

\title{\bf Efficient Postprocessing Procedure for Evaluating Hamiltonian Expectation Values in Variational Quantum Eigensolver.
}

\usepackage{authblk}

\author[1,2,*]{Chi-Chun Chen\thanks{* r08222060@ntu.edu.tw}}
\author[1,2,3,$\dagger$]{Hsi-Sheng Goan\thanks{$\dagger$ goan@phys.ntu.edu.tw}}
\affil[1]{Physics Division, National Center for Theoretical Sciences, Taipei 106319, Taiwan}
\affil[2]{Center for Quantum Science and Engineering, National Taiwan University, Taipei 106319, Taiwan}
\affil[3]{Department of Physics, National Taiwan University, Taipei 106319, Taiwan}

\begin{document}

\maketitle
\thispagestyle{plain}
\pagestyle{plain}

\begin{abstract}
We proposed a simple strategy to improve the postprocessing overhead of evaluating Hamiltonian expectation values in Variational quantum eigensolvers (VQEs). Observing the fact that for a mutually commuting observable group G in a given Hamiltonian, $\expval{G}{b}$ is fixed for a measurement outcome bit string $b$ in the corresponding basis, we create a measurement memory (MM) dictionary for every commuting operator group G in a Hamiltonian. Once a measurement outcome bit string $b$ appears, we store $b$ and $\expval{G}{b}$ as key and value, and the next time the same bit string appears, we can find $\expval{G}{b}$ from the memory, rather than evaluate it once again. We further analyze the complexity of MM and compare it with commonly employed post-processing procedure, finding that MM is always more efficient in terms of time complexity.
We implement this procedure on the task of minimizing a fully connected Ising Hamiltonians up to 20 qubits, and H\textsubscript{2}, H\textsubscript{4}, LiH, and H\textsubscript{2}O molecular Hamiltonians with different grouping methods. For Ising Hamiltonian, where all $O(N^2)$ terms commute, our method offers an $O(N^2)$ speedup in terms of the percentage of time saved. In the case of molecular Hamiltonians, we achieved over $O(N)$ percentage time saved, depending on the grouping method.


\end{abstract}

\begin{keywords}

Quantum computing, Variational quantum eigensolver, VQE, Measurement memory

\end{keywords}

\section{INTRODUCTION}

Variational quantum eigensolvers (VQEs)\cite{mcclean2016theory,kandala2017hardware,cerezo2021variational} have been considered one of the main applications of quantum computers in the noisy intermediate-scale quantum (NISQ) era\cite{preskill2018quantum}. VQEs can be applied to a broad type of tasks, such as chemistry, optimization, and machine learning. A common goal of most of these applications is to minimize an objective function, which usually is represented by a Hamiltonian that can be decomposed into a sum of tensor products of Pauli operators, i.e. 
\begin{equation}
    H = \sum_{i}h_iO_i,
\end{equation}
where $O_i \in \{I,\sigma_X,\sigma_Y,\sigma_Z\}^{\otimes N}$, and $N$ is the number of qubits. 
In VQE, the goal is to minimize the Hamiltonian expectation value, obtained from a statistical average over a large number of measurements. The general process of VQEs can be summarized as follows:

\begin{enumerate}
    \item Design the parameterized ansatz \(U(\mathbf{\theta})\).
    \item Choose the initial parameter vector \(\mathbf{\theta}\).
    \item Calculate the gradient of \(f(\mathbf{\theta}) = \expval{U^\dagger(\mathbf{\theta}) H U(\mathbf{\theta})}{i}\), where \(\ket{i}\) is the initial state.
    \item Update \(\mathbf{\theta}\) according to the gradient, and check if \(f(\theta)\) has decreased.
    \item Go back to step 3 until \(f(\theta)\) cannot be minimized anymore or has reached the maximum iterations.
\end{enumerate}

Assuming choosing a gradient-based classical optimizer, the gradient can be calculated analytically with the parameter shift rule \cite{mitarai2018quantum} for a quantum circuit. However, this process also requires evaluating the expectation value $2$ times at $f(\theta+\frac{\pi}{2}\hat{\theta_x})$ and $f(\theta-\frac{\pi}{2}\hat{\theta_x})$ for each dimension $\theta_x$ in $\mathbf{\theta}$. This results in $2d$ times of evaluations of Hamiltonian expectation value for step 3, where $d$ is the dimension of $\mathbf{\theta}$. Then we evaluate $f(\mathbf{\theta})$ once again with the updated $\mathbf{\theta}$ in step 4, making it a total of $2d+1$ times of evaluation in every optimization step, and thus $(2d+1)T$ evaluations are required for a whole VQE process with $T$ iterations. This number increases as the problem size grows since the number of parameters should grow, and the number of iterations may also grow. It is well known that evaluating a Hamiltonian expectation value once already requires a large number of circuit repetitions \cite{wecker2015progress}. However, as stated above, we even need to repeat this task over and over again to complete the whole VQE process.

There have been many elaborated methods proposed to reduce the circuit repetition for evaluating the Hamiltonian expectation value, such as efficient state tomography \cite{aaronson2019shadow, huang2020predicting, huang2021efficient}, measurement distribution \cite{wecker2015progress,rubin2018application, yen2023deterministic}, and Hamiltonian partitioning \cite{mcclean2016theory, jena2019pauli, verteletskyi2020measurement, yen2020measuring, gokhale2020n, yen2023deterministic}...etc. However, the measurement needed to reach a specific error is still very large. For example, it still requires hundreds of millions of circuit repetitions to evaluate even small molecules \cite{yen2023deterministic} (e.g., H\textsubscript{2}O, NH\textsubscript{3}) to chemical accuracy. Moreover, with such a large number of measurements, the classical overhead of postprocessing becomes nontrivial.

First, we started from a Hamiltonian that is partitioned into commuting groups, i.e.,
\begin{equation}
\label{Hgroup}
    H = \sum_k G_k.
\end{equation}
Since commuting operators share common eigenstates, the Pauli operators in the same group can be evaluated with the same set of measurement outcomes $\{b\}$ in the corresponding measurement basis. The classical cost of evaluating a bit string depends on the number of terms in the group. For example, a fully connected Ising Hamiltonian has all of its operators in the same group, since they are all composed of Pauli Z, and thus requires $O(N^2)$ scaling of cost. A molecular Hamiltonian is known to have $O(N^4)$ operators, and a grouping strategy with $O(N^3)$ groups will thus result in linear scaling of operators in each group, which requires $O(N)$ cost to evaluate. Observing the fact that $\expval{G}{b}$ is fixed and needs to be calculated every time $b$ occurs, a straightforward strategy is to memorize it. Since looking up a bit string with length $N$ in a dictionary requires only $O(1\times N)$ on average. This is the main idea of measurement memory (MM). 

We implemented MM on fully connected Ising Hamiltonians up to 20 qubits, finding an $O(N^2)$ speedup in terms of the percentage of time saved. MM is also tested on Molecular Hamiltonian of H\textsubscript{2}, H\textsubscript{4}, LiH, and H\textsubscript{2}O, each with 3 different grouping strategies: qubit-wise commuting (QWC) grouping \cite{mcclean2016theory, verteletskyi2020measurement}, general commuting (GC) grouping \cite{yen2020measuring}, and Fermion grouping (FG) \cite{huggins2021efficient}. Each grouping method has different scaling of groups and operators in each group, thus leading to a different degree of improvement.



\section{Measurement Memory}
For a Hamiltonian \(H\) that has been divided into \(K\) commuting operator groups, i.e., 
\begin{equation}
\begin{aligned}
    H &= \sum_{k=1}^K G_k, \\
    G_k &= \sum_j h_j^k O_j^k,
\end{aligned}
\end{equation}
where \(O_j^k \in \{I, \sigma_X, \sigma_Y, \sigma_Z\}^{\otimes N}\), and
\begin{equation}
    [O_i^k, O_j^k] = 0.
\end{equation}

The operators in the same commuting groups are simultaneously diagonalizable and can be evaluated with the same set of measurements in the corresponding basis. Measurement Memory (MM) creates an independent dictionary \(\mathcal{M}_k\) for each group \(G_k\) at the beginning. During the future process, every time we evaluate the Hamiltonian expectation value via quantum computer measurement results (including calculating the gradient with the parameter shift rule), we try to find the value \(\expval{G_k}{b}\) in the corresponding dictionary \(\mathcal{M}_k\) for each group with key \(b\). If it doesn't exist, we evaluate it as the normal process and store \(b\) and \(\expval{G_k}{b}\) in the dictionary as key and value.

The pseudocode shown in Algorithm \ref{alg:MM} represents the process of evaluating \(\expval{H}\) once. For each measurement we evaluate, \(\mathcal{M}\) accumulates and starts to develop potential to reduce more and more computational cost.

\begin{algorithm}
\SetKwProg{Try}{try}{:}{}
\SetKwProg{Except}{except}{:}{}
\caption{Measurement Memory}\label{alg:MM}
\(E=0\)

\ForEach{\(G_k\) in \(\{G_1, G_2, \ldots, G_K\}\)}{
    \(E_k = 0\) \\
    \ForEach{\(b\) in \(\{b_1, b_2, \ldots, b_m\}_k\)}{
        \Try{}{
            \(E_k += \mathcal{M}_k[b] \cdot \frac{1}{m}\)
        }
        \Except{}{
            \(\mathcal{M}_k[b] = \expval{G_k}{b}\)

            \(E_k += \mathcal{M}_k[b] \cdot \frac{1}{m}\)
        }
    }
    \(E += E_k\)
}
\Return{\(E\)}
\end{algorithm}

Note that one does not have to sort the measured bit string \(\{b_1, b_2, \ldots, b_m\}\) into probability or count dictionaries in advance, as we usually do in regular procedures when applying MM. We will discuss the complexity of sorting and MM further in the next chapter. We also analyzed the time complexity in the appendix, showing that the potential performance improvement of MM is mainly determined by the scaling of the number of operators in each group. Thus, the improvement will be very limited for an MM created for a sublinear scaling operator group. On the other hand, we should expect fruitful performance improvement for superlinear scaling operator groups.



\section{Comparison with Regular Procedure}
For every commuting group \(G_k\) in a Hamiltonian, we set up the circuit in the corresponding measurement basis and repeat the circuit, say \(m\) times, to obtain a collection of measured bit strings \(\{b_1, b_2, \ldots, b_m\}_k\). Typically, one would sort the set into distinct terms with corresponding counts or probabilities, such as \(\{b_1: Pr(b_1), b_2: Pr(b_2), \ldots, b_L: Pr(b_L)\}_k\), where \(L \leq m\). For simplicity, we refer to the process of evaluating \(\expval{G}\) from \(\{b_1: Pr(b_1), b_2: Pr(b_2), \ldots, b_L: Pr(b_L)\}_k\) as the "sort-and-evaluate" procedure and the process of directly evaluating every bit string in \(\{b_1, b_2, \ldots, b_m\}_k\) as the "naive evaluation". The sort-and-evaluate procedure reduces the evaluation of \(\expval{G}{b}\) by \(m-L\) times. However, the sorting process is not free and might waste a lot of time in the worst case (i.e., \(L=m\)). Nevertheless, we showed in the Appendix that in general and practical cases, one will still prefer the sort-and-evaluate procedure over naive evaluation.

One thing to mention here is that the MM procedure has already incorporated the process of sorting raw measurement results into probability dictionaries. Thus, we do not have to sort the measured result in advance when applying MM. The main difference is that we are now storing their eigenvalues (\(\expval{G}{b}\)) instead of probabilities. While a probability dictionary is useless except for one specific evaluation of the expectation value, the information of MM is able to sustain through the whole VQE and develop the potential to save more and more cost as the iteration goes on. We further proved that the computational cost of the worst-case scenario of MM, which is the case that no eigenvalue of measured bit strings is stored in memory, is identical to the sort-and-evaluate procedure. This indicates that there are no downside trade-offs for adopting MM in terms of time complexity.

\section{Neumerical Simulation}
\subsection{Ising Hamiltonian}
Here we implement MM on fully connected Ising Hamiltonians up to 20 qubits. A random fully connected Ising Hamiltonian can be written as 
\begin{equation}
    \label{IsingHamiltonian}
    H_I = \sum^N_{i=1}\sum^N_{j>i}h_{ij}Z_iZ_j + \sum^N_{i=1} h_iZ_i,
\end{equation}
where $N$ is the qubit number and $h_{ij} \neq 0$. These types of Hamiltonians are commonly seen for solving quadratic unconstrained binary optimization (QUBO) problems~\cite{glover2018tutorial}. The operators all consist of Pauli $Z$, thus we have $\binom{N}{2}+N$ local operator terms in the same commuting group and require only a single measurement basis $\{Z\}^{\otimes N}$.

We observed a special property of the Ising Hamiltonian, given the measured bit string set $\{b\}$,
\begin{equation}
\begin{aligned}
    \expval{H_I} &=\sum_{i}\sum_{j>i}\left[\frac{1}{m}\sum_{b\in \{b\}}h_{ij}\expval{Z_iZ_j}{b}\right]  + \sum_{i}\left[\frac{1}{m}\sum_{b\in \{b\}}h_{i}\expval{Z_i}{b}\right] \\
    &= \frac{1}{m}\sum_{b\in \{b\}}\left[\sum_{i}\sum_{j>i}h_{ij}\expval{Z_iZ_j}{b} + \sum_{i}h_{i}\expval{Z_i}{b}\right] \\
    &= \frac{1}{m}\sum_{b\in \{b\}}\expval{H_I}{b}.
\end{aligned}
\end{equation}
This implies that the cost of post-measurement classical computation per measurement is equivalent to searching one eigenstate. Due to the large number of circuit repetitions for evaluating the Hamiltonian expectation value, it would be very likely that the circuit repetition exceeds the search space of the problem ($2^N$) for problem sizes that are not large enough. It is thus difficult to justify the usage of VQE for Ising Hamiltonian on intermediate-scale quantum computers today and in the near future. MM here provides a simple strategy to bypass this limitation.

The measurement is done after a one CNOT layer ansatz, with $2N$ parameters. We adopted $O(N^2)$ circuit repetition scaling with problem size for each evaluation of Hamiltonian expectation value, proportional to the scaling of operator terms. We take 10 initial parameter guesses, and run VQE for 200 iterations. The final result is an average over 10 incidents. Note that the Hamiltonian is fixed for both normal procedure and MM procedure at every qubit size, and so are those 10 initial guesses.

\begin{center}
\begin{figure*}[h]
    \includegraphics[width=0.95\linewidth]{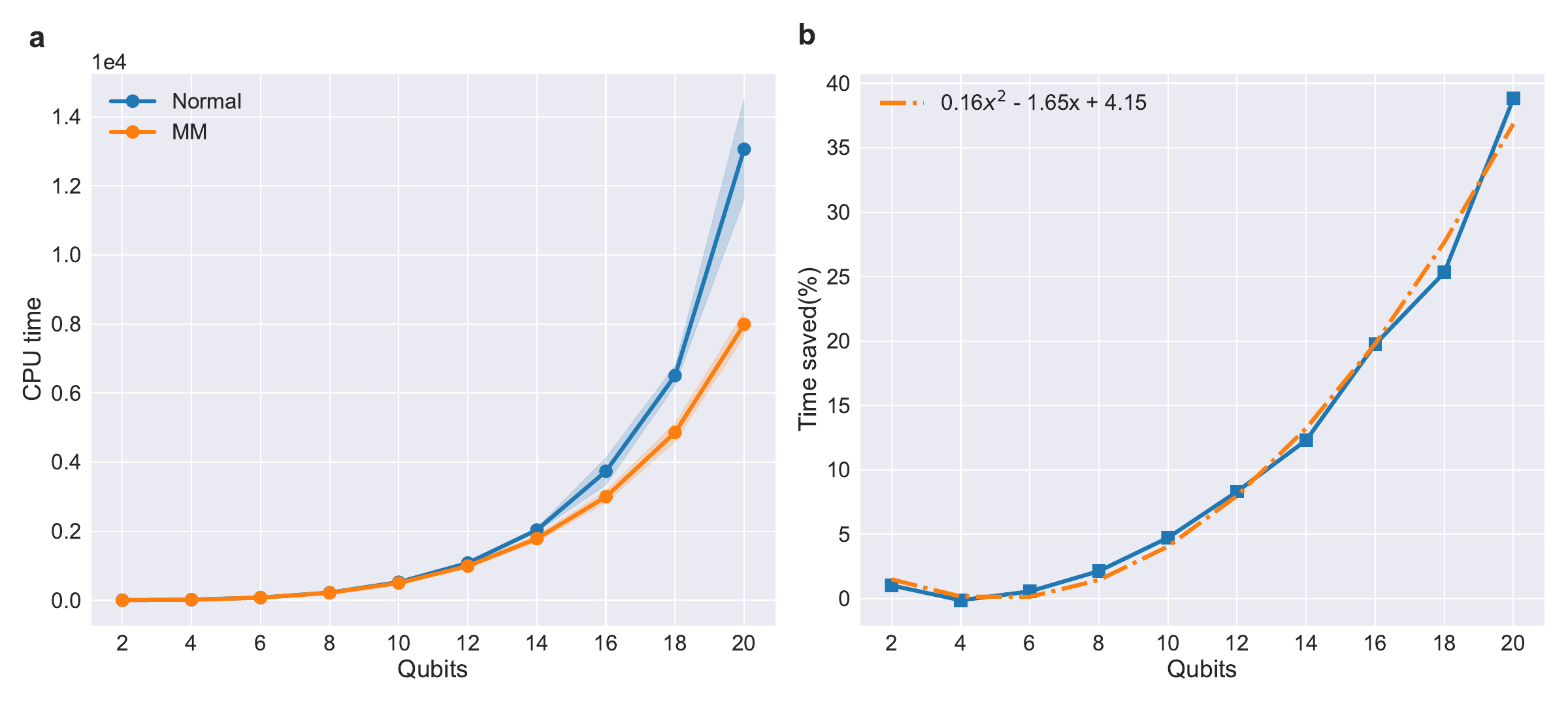}
\caption{\textbf{a} CPU time ($t$) for 200 steps of optimization of fully-connected Ising Hamiltonians with increasing problem size. \textbf{b} Percentage time saved, i.e. $(t_{Normal}-t_{MM}) / t_{Normal}$, with MM.}
\label{fig:Ising_result}
\end{figure*}
\end{center}

The results are shown in Fig \ref{fig:Ising_result}, MM achieved quadratic speed up in terms of percentage compared with the normal evaluation procedure. There is an additional benefit from using MM for optimization problems. Since the whole Ising Hamiltonian is one commuting group, $\expval{G}{b} = \expval{H_I}{b}$, where $b$ is the measured bit string and also an eigenstate of $H_I$. Thus, we are able to store every eigenvalue of every eigenstate that ever measured during the whole VQE process. This largely increases the probability of finding a good solution (i.e. low energy eigenstate) due to the large number of measurements.

\input{table/grouping}

\subsection{Molecular Hamiltonian}
It is well known that the Pauli operator terms scales $O(N^4)$ with system size for molecular Hamiltonian. Fortunately, several method have been proposed to partition those operators into commuting groups\cite{verteletskyi2020measurement, yen2020measuring, huggins2021efficient}. Although it may not directly reduce the circuit repetition without gingerly selection of grouping terms\cite{huggins2021efficient}, MM is able to reduce more classical overhead of postprocessing the more terms are grouped together. We test MM on three grouping strategies, qubit-wise commuting (QWC) grouping\cite{mcclean2016theory, verteletskyi2020measurement}, general commuting (GC) grouping\cite{yen2020measuring}, and Fermion grouping (FG)\cite{huggins2021efficient}. QWC results in $O(N^4)$ scaling with constant scaling terms in each group, while GC results in $O(N^3)$ scaling and linear scaling group members. FG is a more special case since it permits reasonable discarding of small eigenvalues in second quantization Hamiltonian, resulting in $O(N)$ scaling groups, where each group contains $O(N^2)$ terms for small size molecules and reaching $O(log^2(N))$ as the system size become large\cite{motta2021low}. 

\begin{center}
\begin{figure*}[h]
    \includegraphics[width=\linewidth]{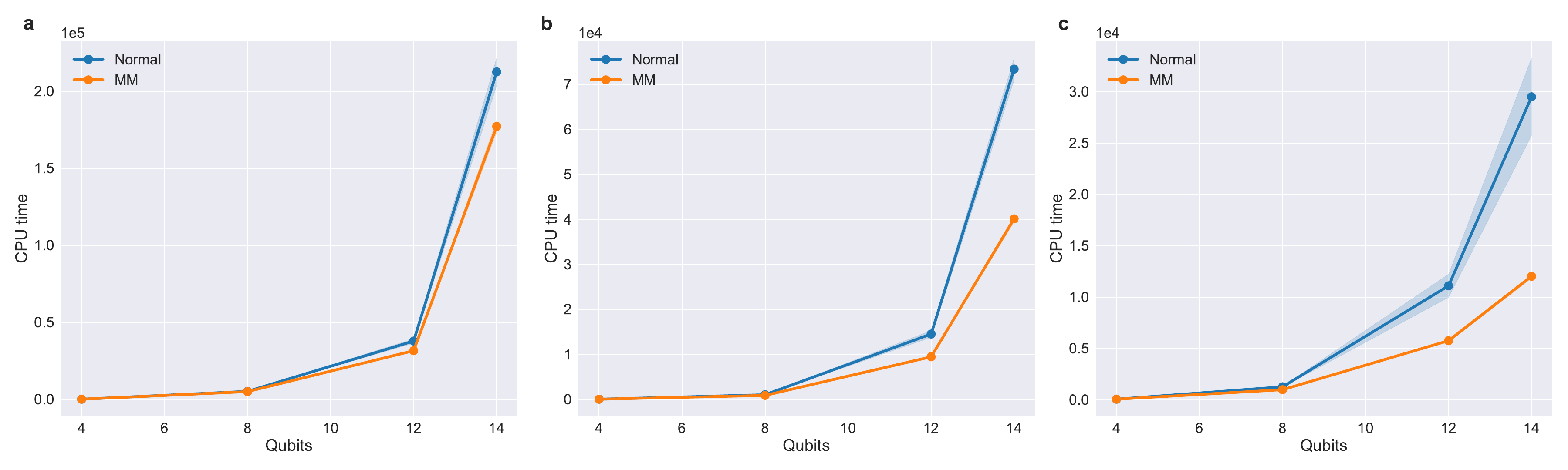}
\caption{CPU time required for 100 steps of optimization, with different grouping methods and $N$ scaling of circuit repetition for each commuting group with qubit. \textbf{a} QWC, \textbf{b} GC, \textbf{c} FG, of H\textsubscript{2} (4 qubit), H\textsubscript{4} (8 qubit), LiH (12 qubit), and H\textsubscript{2}O (14 qubit) molecular Hamiltonians with a minimum basis set. The shaded regions are the standard deviation of 10 determined initial parameter guesses.}
\label{fig:mol_result}
\end{figure*}
\end{center}

\begin{center}
\begin{figure}[h]
    \includegraphics[width=0.95\linewidth]{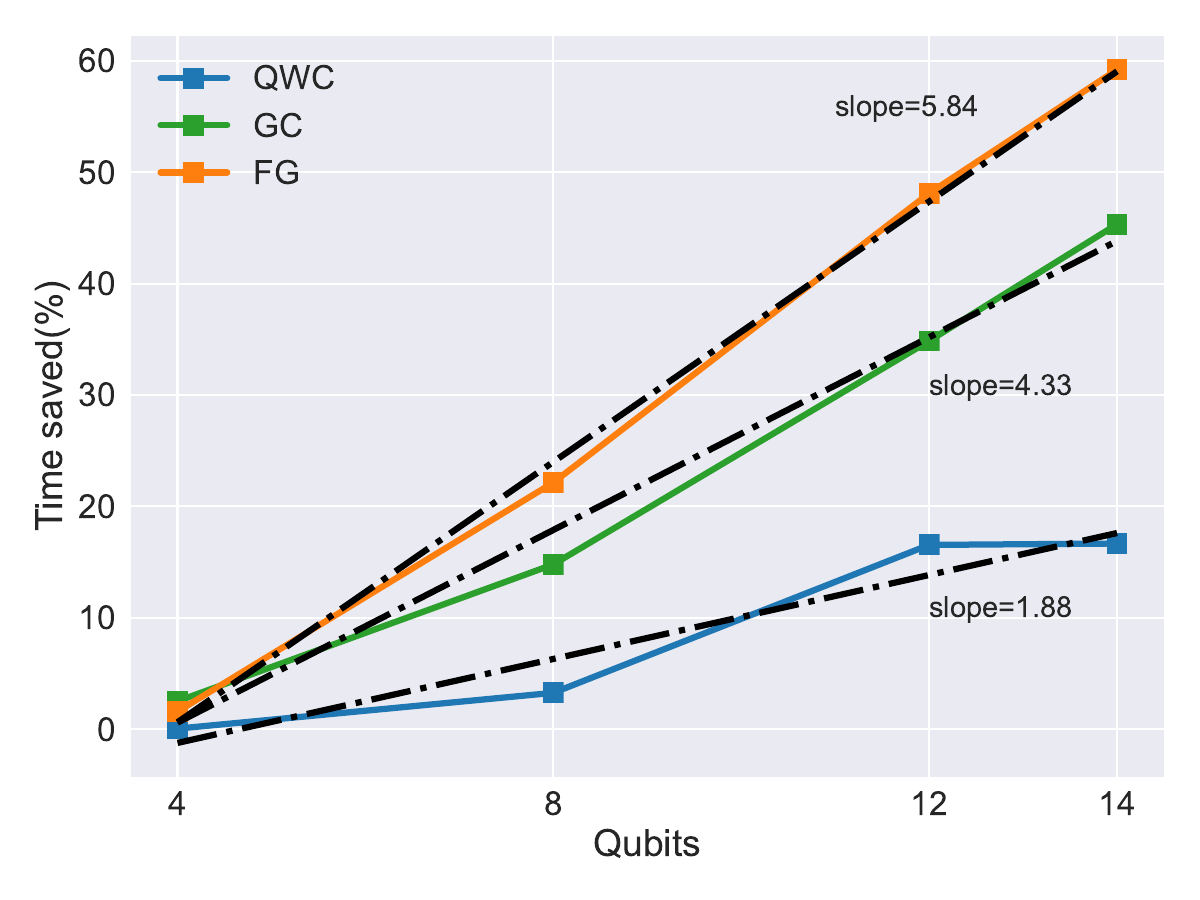}
\caption{Percentage time saved, i.e.  $(t_{Normal}-t_{MM}) / t_{Normal}$, with MM, using different grouping methods for molecular Hamiltonians.}
\label{fig:Mol_timesaved}
\end{figure}
\end{center}

Here we demonstrated the improvement of MM on H\textsubscript{2} (4 qubit), H\textsubscript{4} (8 qubit), LiH (12 qubit), and H\textsubscript{2}O (14 qubit) molecules with sto-3g basis set, transformed via Jordan–Wigner (J-W) transformation. The comparison of MM with normal evaluation procedure is shown in Fig \ref{fig:mol_result}. As expected, the order of improvement is FG > GC > QWC, since the improvement of MM is more significant if there are more terms in the same group, and also because FG results in more terms of Pauli operators in total (TABLE \ref{table:grouping}). The time saved in terms of percentage is also shown in Fig \ref{fig:Mol_timesaved}. All three grouping methods (i.e. QWC , GC, and FG grouping) are able to achieve linear to superlinear improvement in percentage time saved. However, for FG case, the qubit sizes we are able to demonstrate here are at the transition point of the term scaling, i.e. $O(N^2)$ to $O(log^2(N))$, making it more difficult to estimate the scaling of improvement of MM for larger qubit systems.


\section{CONCLUSIONS and OUTLOOK}
In this article, we introduced a Measurement Memory (MM) dictionary designed to store a measured bit string $b$ and its eigenvalue $\expval{G}{b}$ of a commuting group in a Hamiltonian. Throughout the VQE process, MM accumulates records, reducing the computational cost of evaluating $b$ over all terms in $G$ each time $b$ occurs after the first instance. We achieved $O(N^2)$ percentage time savings for a fully connected Ising Hamiltonian, providing an additional benefit of increasing the probability of finding a low-energy solution. MM also provides over $O(N)$ time savings for molecular Hamiltonian depending on the grouping method. With careful selection of grouping elements, one may be able to reduce both circuit repetition and postprocessing classical overhead.
We further compare the time complexity of regular sort-and-evaluate procedure with MM, finding that that MM is a more efficient way of storing useful information for evaluating Hamiltonian expectation values in VQE, which sustains through the whole process. Moreover, the worst case scenario of MM is identical to the regular procedure. Thus, MM is a more efficient procedure of doing post-processing for evaluating Hamiltonian expectation value in VQE, and employing MM has no down side trade offs.
For future applications, MM can also be applied to other deterministic measurement schemes, such as other grouping methods, and derandomized shadow\cite{huang2021efficient} as long as the measurement basis are fixed.

\bibliographystyle{ieeetr}
\bibliography{ref}

\newpage

\onecolumn
\appendix
\label{appendixA}

\noindent\textbf{Time Complexity of sort-and-evaluate}\newline
Given $\{b_1, b_2, ... b_m\}$ with m total terms and $L$ distinct term, which can be sorted into dictionary $\{b_1:Pr(b_1), b_2:Pr(b_2), ... b_L:Pr(b_L)\}$, where $L\leq m$. The sorting process is 
\begin{algorithm*}
\SetKwProg{Try}{try}{:}{}
\SetKwProg{Except}{except}{:}{}
\caption{sort measurement}\label{alg:sort}
$D$ = \{\}\\
\For{$b$ in $\{b_1, b_2, ... b_m\}$}{
    \If{$b$ in $D$}{
        $D[b]$ = $D[b]$ + $\frac{1}{m}$\\
    }
    \Else{
        $D[b]$ = 1\\
    }
    }
\end{algorithm*}

First we analyze the complexity for sort-and-evaluate method. Dictionaries in python are implemented with hash table, checking if a new key $b$ is in $D$ requires calculating the hash function, which has cost $\mathcal{O}(N)$ since each $b$ has length $N$. If hash function points to an empty memory, we know $b$ is not in $D$, then we add it to $D$ with $\mathcal{O}(1)$. If hash function points to an occupied memory, we need to check if $b$ is the same bit string as the occupied one to prevent from hash collision, costing additional $\mathcal{O}(N)$. With the bit string set given above, we would result in $L$ "no"s and $m-L$ "yes"s (whether the memory is occupied). Since we only have to evaluate $\expval{G}{b}$ for distinct $b$s, making the complexity for sort-and-evaluate 
\begin{equation}
    L\left[\mathcal{O}(N)+\mathcal{O}(G)\right]+(m-L)2\mathcal{O}(N)),
\label{s&e}
\end{equation}
where $\mathcal{O}(G)$ is the complexity of evaluating $\expval{G}{b}$ for one bit string.

On the other hand, naive evaluation simply evaluates $\expval{G}{b}$ for every $b$ in $\{b_1, b_2, ... b_m\}$, making the complexity
\begin{equation}
    m\mathcal{O}(G).
\label{naive_e}
\end{equation}
The condition for choosing sort-and-evaluate over naive evaluation is 
\begin{equation}
    L\left[\mathcal{O}(N)+\mathcal{O}(G)\right]+(m-L)2\mathcal{O}(N)) \leq m\mathcal{O}(G), 
\label{s&econdition}
\end{equation}
and thus 
\begin{equation}
    \mathcal{O}(G) \geq \frac{2m-L}{m-L}\mathcal{O}(N).
    \label{s&econdition2}
\end{equation}
Now we consider two extreme cases, first is an ansatz with same probability distribution over all eigenstates. Sampling from this ansatz is equivalent to a random sample. In this case, we can estimate $L$ by calculating the expectation value of distinct values of drawing $m$ times randomly from the $2^N$ binary string space,
\begin{equation}
    \mathbb{E}[L] = 2^N\left(1- \left(1-\frac{1}{2^N}\right)^m\right),
\end{equation}
and $\mathbb{E}[L] \rightarrow m$ as $2^N \rightarrow \infty$.
Thus, substituting $L$ for $m$ in condition \ref{s&econdition2}, we end up with $O(G) \geq \infty$, which implies that one should conduct naive evaluation in this case. 
However, despite the fact that naive evaluation is always better in this case, the degree of improvement may not be significant. By comparing the complexity of two schemes with $L \rightarrow m$, the $m\mathcal{O}(N)$ improvement, although may be large if $m$ scales badly, is not the bottleneck if $\mathcal{O}(G) \geq \mathcal{O}(N)$. Since evaluating $\expval{G}{b}$ requires looping through every Pauli word on each qubit for every operator in the group, 
\begin{equation}
    \mathcal{O}(G) = \mathcal{O}(N)\mathcal{O}(T),
\end{equation}
where $\mathcal{O}(T)$ is the scaling of operator terms in group $G$. This shows that $\mathcal{O}(G) \geq \mathcal{O}(N)$ is always true and thus the additional cost of adapting sort-and-evaluate scheme will not be the bottleneck. 
The second case is a highly concentrated ansatz, i.e. $L << m$. In this case, condition \ref{s&econdition2} becomes 
\begin{equation}
    \mathcal{O}(G) \geq \mathcal{O}(2N), 
\end{equation}
suggesting that if the complexity of evaluating $\expval{G}{b}$ is worse than linear, which, as mentioned above, is true for general cases, one should conduct the sort-and-evaluate scheme.  

This discussion thus conclude that one can always conduct the sort-and-evaluate scheme with advantage in most case, and with acceptable disadvantage in some cases.
\\

\noindent\textbf{Time Complexity of MM}\newline
As shown in algorithm \ref{alg:MM}, for every bit string in $\{b_1, b_2, ... b_m\}$, we check if $b$ is in $\mathcal{M}$. It also takes $O(N)$ to calculate the hash function to see if the corresponding memory is occupied. A "no" requires evaluating $\expval{G}{b}$, and a "yes" requires checking if it is exact the same string. 
Assuming the same set of bit string mentioned in the first part of Appendix, where we have total $m$ bit strings with $L$ distinct ones. We also have $\ell$ bit strings in $L$ distinct bit strings stored in $\mathcal{M}$ with its corresponding eigenvalue $\expval{G}{b}$. 
We will thus get $L-\ell$ "no"s and $m-L+\ell$ "yes"s, and the complexity for naive evaluation of $\expval{G}$ given  $\mathcal{M}$ is 
\begin{equation}
    (L-\ell)\left[\mathcal{O}(N)+\mathcal{O}(G)\right]+(m-L+\ell)2\mathcal{O}(N)).
\label{complexMM}
\end{equation}

Now, if one sort the bit string into probability dictionary before feeding into MM, i.e. the $\{b_1, b_2, ... b_m\}_k$ in the $4^{th}$ step in algorithm \ref{alg:MM} becomes $\{b_1:Pr(b_1), b_2:Pr(b_2), ... b_L:Pr(b_L)\}$, and all following $\frac{1}{m}$ becomes $Pr(b)$. The complexity is thus 
\begin{equation}
    L\mathcal{O}(N)+(m-L)2\mathcal{O}(N) + (L-\ell)\left[\mathcal{O}(N)+\mathcal{O}(G)\right] + \ell2\mathcal{O}(N).
\label{complexMM_sorted}
\end{equation}
The first two terms represent the cost for sorting, and the last two terms represent the cost of finding the sorted $L$ bit strings in $\mathcal{M}$. By comparing equation \ref{complexMM} and equation \ref{complexMM_sorted}, we see that sorting before MM introduced additional $\left(L+2\ell\right)\mathcal{O}(N)$ computational cost. Although it is also not the bottleneck in most cases, it is totally unnecessary to sort in advance when applying MM.

In equation \ref{complexMM}, we can see that in the worst case, for example, the first step, $\ell=0$, the complexity is equal to sort-and-evaluate scheme, i.e. equation \ref{s&e}. For every extra state we store (i.e. $\ell+1$), we are trading $\mathcal{O}(N)+\mathcal{O}(G)$ cost for $2\mathcal{O}(N)$. Since $\mathcal{O}(G) \geq \mathcal{O}(N)$ in general (one will have to evaluate through the pauli word on every qubit, including "I"), we explicitly proofed that MM is more efficient than the original sort-and-evaluate procedure (without MM). Furthermore, if given enough iteration, as $\mathcal{M}$ accumulates, we expect asymptotic behavior of $L-\ell$ to zero, making the complexity 
\begin{equation}
    m\mathcal{O}(2N).
\label{MM_manyitr}
\end{equation}

\end{document}

%% file: table/grouping.tex
\begin{table}
\begin{center}
\begin{tabular}{ |c | c | c| c | c| c | c| }
\hline
     Mol  &  \multicolumn{2}{c|}{QWC} & \multicolumn{2}{c|}{GC} & \multicolumn{2}{c|}{FG}  \\
\hline
     & total & groups & total & groups & total & groups\\
\hline
    H\textsubscript{2} & 15 & 5 & 15 & 2 & 34 & 4\\
\hline
    H\textsubscript{4} & 185 & 67 & 185 & 9 & 317 & 11\\
\hline
    LiH & 631 & 151 & 631 & 34 & 877 & 22\\
\hline
    H\textsubscript{2}O & 1086 & 556 & 1086 & 90 & 1611 & 29\\
\hline
    
\end{tabular}
\caption{Total operator terms and commuting groups for different grouping methods.}
\label{table:grouping}
\end{center}
\end{table}